# The SST Fully-Synchronous Multi-GHz Analog Waveform Recorder with Nyquist-Rate Bandwidth and Flexible Trigger Capabilities

Stuart A. Kleinfelder,[1*] *Senior Member, IEEE*, Edwin Chiem[1] and Tarun Prakash[1]

*Abstract—* The design and performance of a fully-synchronous multi-GHz analog transient waveform recorder I.C. ("SST") with fast and flexible trigger capabilities is presented. The SST's objective is to provide multi-GHz sample rates with intrinsically-stable timing, Nyquist-rate sampling and high trigger bandwidth, wide dynamic range and simple operation. Containing 4 channels of 256 samples per channel, the SST is fabricated in an inexpensive 0.25 micrometer CMOS process and uses a high-performance package that is 8 mm on a side. It has a 1.9V input range on a 2.5V supply, exceeds 12 bits of dynamic range, and uses ~128 mW while operating at 2 G-samples/s and full trigger rates. With a standard 50 Ohm input source, the SST exceeds ~1.5 GHz -3 dB bandwidth. The SST's internal sample clocks are generated synchronously via a shift register driven by an external LVDS oscillator running at half the sample rate (e.g., a 1 GHz oscillator yields 2 G-samples/s). Because of its purely-digital synchronous nature, the SST has ps-level timing uniformity that is independent of sample frequencies spanning over 6 orders of magnitude: from under 2 kHz to over 2 GHz. Only three active control lines are necessary for operation: Reset, Start/Stop and Read-Clock. When operating as common-stop device, the time of the stop, modulo 256 relative to the start, is read out along with the sampled signal values. Each of the four channels integrates dual-threshold trigger circuitry with windowed coincidence features. Channels can discriminate signals with ~1mV RMS resolution at >600 MHz bandwidth. Comparator thresholds can be set individually, and their outputs are directly available for simple threshold monitoring and rate control. Alternatively, their High and Low results can be AND'd over an adjustable window of time in order to exclusively trigger on bipolar impulsive signals. Trigger outputs can be CMOS or low-voltage differential signals, e.g. positive-ECL (0-0.8V) for low-noise operation.

## I. INTRODUCTION

ANALOG waveform sampling and digitization circuits have seen wide use in applications that require large-scale, low-power transient signal acquisition. Early versions of these "SCA" circuits, developed in 1988-1990, e.g. [1], provided 13 bits of dynamic range with 100 MHz sample rates, and were used in numerous Nuclear Science and High-Energy Physics experiments. A subsequent generation from 1992 achieved 5 G-samples/s rates by using an analogically-adjustable digital delay line to generate its high-speed sampling clocks [2]. Numerous derivations included parallel analog-to-digital conversion, e.g. "ATWD" [3], and were widely used in the field of Particle Astrophysics, e.g. [4]. Continued evolution led to on-chip phase-locked-loop-based clock generation and pattern-matching trigger circuitry [5, 6].

This paper describes a new generation of synchronous multi-GHz devices, the first of which is the present 4-channel, 256 cell per channel version. A principle advantage of its fully-synchronous clock over prior asynchronous or PLL-based clocking is that ps-level sample-to-sample timing uniformity is obtained independently of the acquisition clock speed, which can span many orders of magnitude. The new device, dubbed "SST" for Synchronous Sampling plus Triggering, includes optimized design and packaging to reach Nyquist-rate bandwidth (-3 dB point is over 1.5 GHz) along with flexible trigger capabilities that operate at >600 MHz bandwidth and with ~1 mV RMS accuracy.

## II. HIGH-SPEED CLOCK GENERATION

The SST's high speed sampling clocks are generated by a fast shift register containing a single "1" as a pointer. As shown in Fig. 1, the base input clock is provided via an LVDS receiver followed by a 2-phase clock generator, which feeds a buffer tree that distributes the clock in a highly-uniform fashion to the shift register. The shift register is dynamic and consists of 128 master-slave flip-flops configured in a circular fashion. Both the master and slave sections of the flip-flops are used to generate sample clocks, and hence the sample rate is doubled by interleaving, e.g. a 1 GHz clock results in a 2 G-sample/s rate. Being formed exclusively by clocked digital logic, the sampling rate range is 6 orders of magnitude wide (≤2 kHz to ≥2 GHz). Its timing uniformity remains consistent regardless of its clock speed, based most substantially on that of the LVDS clock source, for which ≤1ps jitter is commonly achieved (e.g., Fox "XpressO Ultra" oscillators).

In many applications, the SST would be operated in a common-stop mode, endlessly sampling until it is stopped by a trigger, at which time the preceding samples contain the signal of interest. The position of the stop may hence be random relative to the start, yet knowing its position remains important in order to delineate the beginning and end of the record. Therefore, the position of the pointer at the moment it is stopped is read out in parallel with the analog samples. Unlike many PLL-based schemes, the SST can be started and stopped instantly, and common-start operation is equally easy.

* Corresponding author: Stuart A. Kleinfelder, [1]University of California, 4416 Engineering Hall, Irvine CA, 92697, U.S.A. Voice: 949-824-9430, Email: stuartk@uci.edu.

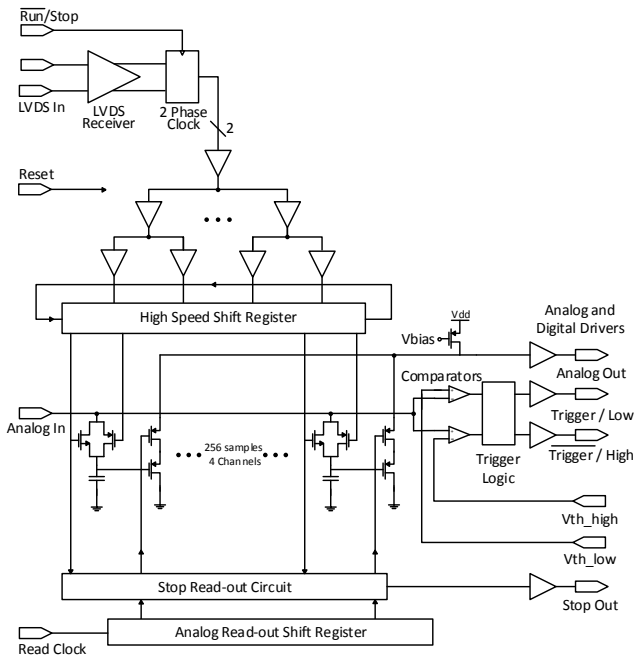

Fig. 1: Device overview showing clock features and one channel of four.

## III. SAMPLING AND READOUT

The SST uses an economical 0.25 μm, 2.5V CMOS process and has a wide input range of 1.9V. Analog sampling is performed by complimentary CMOS switches and 80 fF metal-insulator-metal capacitors, leading to a low sample leakage rate and a measured temporal noise of 0.42 mV, RMS, and hence a dynamic range of 12 bits is achieved. Great care was taken in designing the analog input path to keep its resistance and capacitance low. A small, 8 by 8 mm, fine-pitch surface-mount package was also chosen to keep package-related capacitance and bond-wire inductance low (Fig. 2). The resulting bandwidth using a standard 50 Ohm source is flat to within about +/- 0.5 dB out to ~1.3 GHz, and its -3 dB bandwidth is over 1.5 GHz.

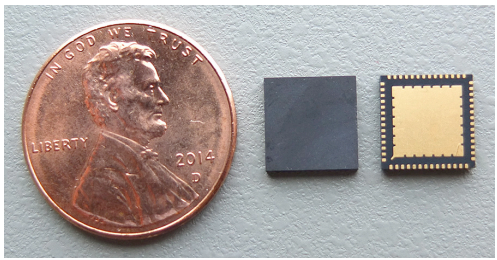

Fig. 2: Device in 8 mm packaging (top and underside shown separately).

Fig. 3 shows a 100 MHz sine wave being captured at 2 G-samples/s. Fig. 4 shows a 100 Hz sine wave being sampled at the extremely low rate of 2 k-samples/s, thus demonstrating 6 orders of magnitude of sample speed range. Sampling at 2kHz is indeed extraordinarily slow, and some leakage appears, especially on a few outlying samples, due to the very long acquisition time (0.128 s), but the SST remains actually usable at these low-audio rate speeds. Average leakage is about 150 mV/s, which is well under the thermal noise of the chip at more typical sample rates of say 1 MHz and above.

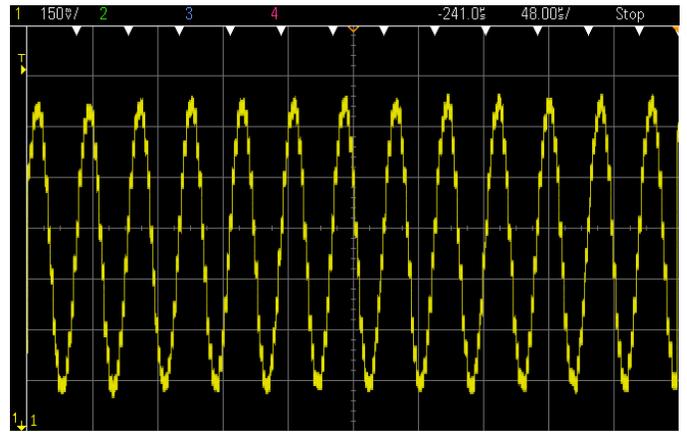

Fig. 3: SST analog readout of a 100 MHz sine wave sampled at 2 GHz.[1]

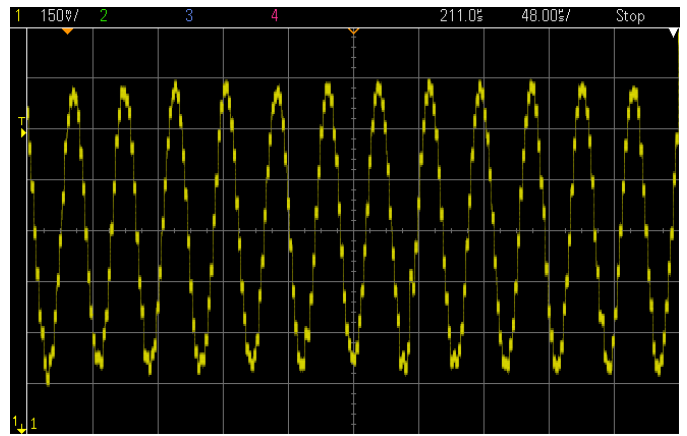

Fig. 4: SST analog readout of a 100 Hz sine wave sampled at 2 kHz.[1]

## IV. TRIGGER GENERATION

The SST incorporates high-performance real-time trigger circuitry, depicted in Fig. 4. Each channel includes two fast comparators, intended for High and Low thresholds. Each pair of comparators observes its channel's analog input directly, with no gain, buffering or level-shifting before-hand. With direct, individual control over comparator inputs, setting thresholds is an easy process. For both high-speed and high gain, the comparators use a large number of fast (~1.2 GHz bandwidth) but low-gain (~3.3 V/V) stages. The comparators can thus discriminate voltage differences with (conservatively) over 600 MHz bandwidth, e.g. small pulses with 500 ps full-width at half-maximum are discriminated, and they achieve a voltage sensitivity of 1 mV, RMS.

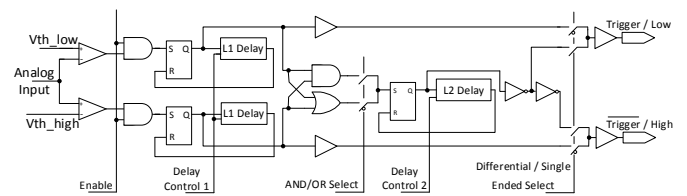

Fig. 4: Trigger logic including inhibit, time-stretching latches and delay lines, AND/OR selection, and selection of dual single-ended or differential outputs.

[1] Raw analog readout without pedestal subtraction. Includes noise at ~89 MHz picked up by the oscilloscope probe from UCI's college FM station.

The results are processed through pulse-stretching latches and analogically-controlled digital delay lines that hold the comparator outputs high for an amount of time that is typically set based on the frequency of interest, e.g. 5 ns when searching for bipolar signals of 100 MHz or higher (~4-100ns). A coincidence circuit then computes the AND or OR of the stretched comparator outputs, and passes this result into a final stretching circuit, capable of stretching pulses from ~10-1,000ns in width. The latter insures that external trigger circuits can respond in time and/or conveniently allows for the external formation of coincidences across channels. The trigger logic permits a net trigger rate of ~250 MHz per comparator or a combined ~100 MHz per channel.

Because the trigger outputs must operate without interfering with the analog sampling, the SST can be configured to generate low-voltage differential trigger outputs, e.g. with positive-ECL levels (0-0.8V) or 1.2V CMOS. As seen in Fig. 5, the trigger output drivers detects whether low-voltage operation is desired by the application of the chosen output voltage power supply ("Output Vdd"), and switches from the use of ordinary P-channel pull-up output FETs to N-channel output FETs, which maintain optimal rise times despite the lower output power supply voltage. As an alternative to a differential output, the pair can be configured as a low-voltage single-ended pair, providing direct access to the individual stretched comparator outputs. Thus, trigger thresholds and rates can easily be monitored on a per-comparator basis.

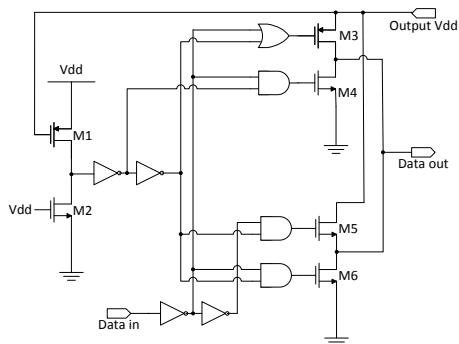

Fig. 5: Trigger output driver with automatic selection of standard CMOS or low-voltage outputs, e.g. 1.2V or 0.8V (i.e., positive ECL).

## V. SUMMARY

A fully synchronous analog transient waveform recording and triggering I.C., the "SST," containing 4 channels of 256 cells per channel and flexible on-chip trigger circuitry is described. It was fabricated in a 0.25μm, 2.5V process to preserve a wide input range (0-1.9V), low sampling noise (~0.42 mV RMS) and low signal leakage for high dynamic range (12 bits). Its typical sample rate is 2 GHz. Optimized design and packaging yielded a nearly flat analog bandwidth to ~1.3 GHz using a standard 50-Ohm signal source, and a -3 dB bandwidth of ~1.5 GHz. Maximum power consumption is 128 mW with all features including triggers operating at full speed. The SST includes a dual-threshold coincidence trigger per channel that operates with ~1 mV RMS resolution at >600 MHz bandwidth and delivers LVDS outputs. The ARIANNA experiment is currently deploying systems based on the SST chip in Antarctica [7]. The SST provides low-cost, flexible, high sampling and trigger performance that covers a very wide range of potential applications.

Table I: SST Figures of Merit

| Parameter | Value |
|---|---|
| Technology: | 0.25 μm CMOS |
| No. of channels and samples/ch: | 4 channels of 256 samples |
| Packaging: | 8 by 8mm 56 pin PLCC |
| Input clock (typical): | 1 GHz LVDS |
| Sample rate (typical): | 2 G-samples/s |
| Minimum sample rate: | 2 k-samples/s |
| Supply voltage: | 2.5V |
| Analog input range: | 0-1.9V |
| Theoretical thermal noise: | ~ 0.25 mV, RMS |
| Input-referred temporal noise: | ~ 0.42 mV, RMS |
| Dynamic range as measured: | > 12 bits, RMS |
| Max. power including triggering: | 128 mW at 2 G-samples/s |
| Analog bandwidth: | > 1.5 GHz, -3dB |
| Average analog leakage rate: | ~ 0.15 V/s |
| Fixed pattern (pedestal) noise: | ~ 6.5 mV, RMS |
| Trigger comparators per channel: | 2 (typ. High and Low) |
| Trigger sensitivity: | < 1 mV, RMS |
| Trigger bandwidth: | > 600 MHz (conservative) |
| Trigger functions per channel: | AND/OR, windowed |
| Trigger output modalities: | Differential/single-ended |
| Trigger output voltage: | 0.8, 1.2 or 2.5V CMOS |

### ACKNOWLEDGMENTS

We thank Wei Cai for her assistance designing the SST's LVDS receiver. We also thank Steve Barwick and numerous other members of the ARIANNA collaboration for invaluable input. This work was supported by funding from the Office of Polar Programs and Physics Division of the U.S. National Science Foundation, grant awards ANT-08339133, NSF-0970175 and NSF-1126672, by NASA 14-NIAC14B-0269, and the Dept. of Physics and Astronomy, Uppsala University.